\documentclass[pra,twocolumn,floatfix,a4paper,superscriptaddress]{revtex4}
\usepackage{bm,color,graphicx,amsmath,txfonts}
\usepackage{here}
\usepackage{array}
\usepackage{graphicx}
\usepackage[colorlinks, citecolor=blue,linkcolor=blue]{hyperref}
\usepackage{braket}
\usepackage{dsfont}
\usepackage{csquotes}
\begin{document}

\makeatletter
\renewcommand{\@biblabel}[1]{\makebox[2em][l]{\textsuperscript{\textcolor{black}{\fontsize{10}{12}\selectfont[#1]}}}}
\makeatother

\let\oldbibliography\thebibliography
\renewcommand{\thebibliography}[1]{%
  \addcontentsline{toc}{section}{\refname}%
  \oldbibliography{#1}%
  \setlength\itemsep{0pt}%
}

\title{Reply to Comment on "Controlling the Dynamical Evolution of Quantum Coherence and Quantum Correlations in $e^{+}e^{-} \to \Lambda\bar{\Lambda}$ Processes at BESIII"}

\author{Elhabib Jaloum}
\affiliation{LPTHE-Department of Physics, Faculty of Sciences, Ibnou Zohr University, Agadir, Morocco}

\author{Mohamed Amazioug}
\email{m.amazioug@uiz.ac.ma}
\affiliation{LPTHE-Department of Physics, Faculty of Sciences, Ibnou Zohr University, Agadir, Morocco}


\begin{abstract}
In the annihilation process $e^+e^- \to J/\psi \to \Lambda\bar{\Lambda}$ via $J/\psi$, the hyperon--antihyperon pair emerges from the same non-perturbative QCD hadronization process, where both spins develop in a common dense partonic environment. We describe the evolution of the $\Lambda\bar{\Lambda}$ spin degrees of freedom using correlated quantum channels that represent the effective influence of the common QCD hadronization environment on the two-spin system. The parameter $\mu$ characterizes the degree of environmental correlation resulting from the common hadronization process. Using the experimentally reconstructed spin density matrix, we evaluate quantum steering and investigate its robustness and dynamical evolution under correlated Markovian and non-Markovian quantum channels. 
\end{abstract}
\maketitle

\begin{figure}[!h]
\begin{center}
\includegraphics[width=9cm,height=6cm]{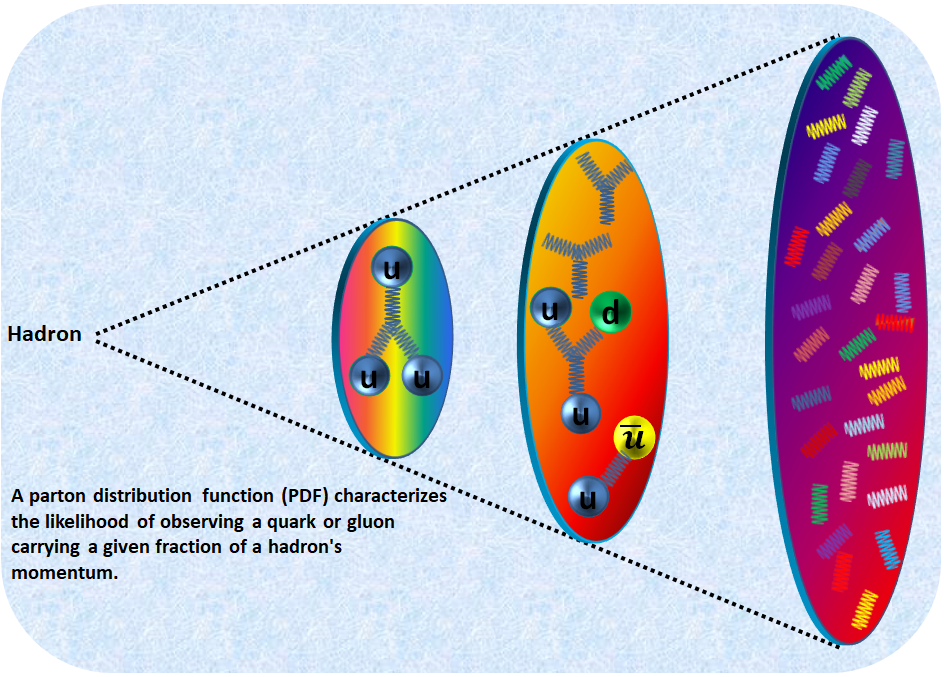}
\includegraphics[width=9cm,height=6cm]{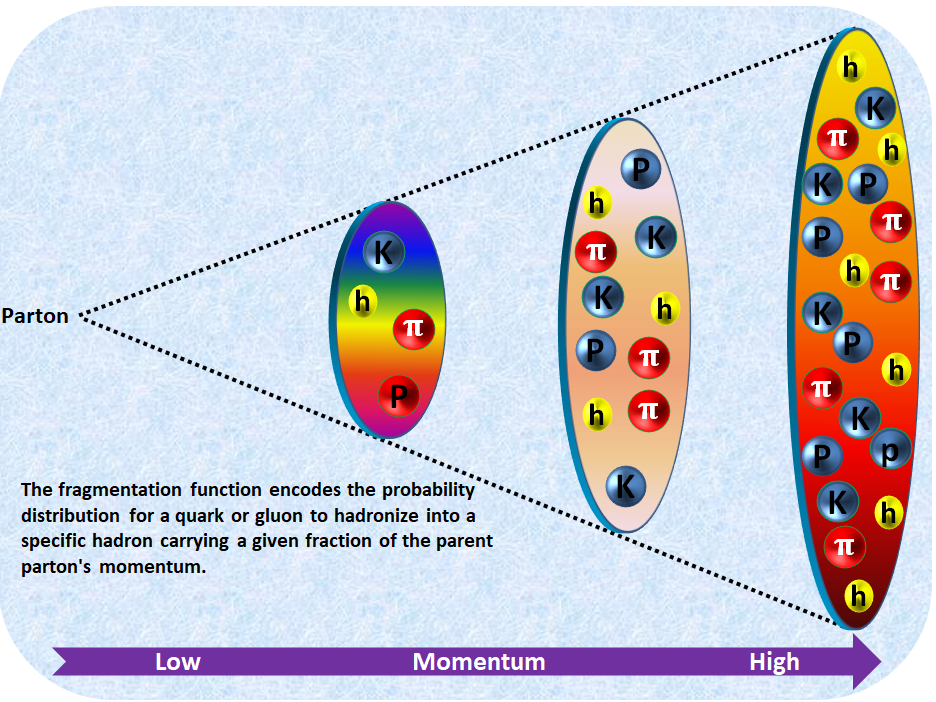}
\end{center}
\caption{Diagram Showing the Hadron–Parton Transition Across Momentum Scales and Their Links via Parton Distribution Functions (PDFs) and Fragmentation Functions (FFs).}
\label{fig:Int}
\end{figure}

The $\Lambda\bar{\Lambda}$ system investigated in Ref.~\cite{JA} is produced in the annihilation process $e^+e^- \to J/\psi \to \Lambda\bar{\Lambda}$. Its formation proceeds through perturbative partonic dynamics followed by non-perturbative hadronization (Fig.~\ref{fig:Int}). Immediately after the decay of the intermediate $J/\psi$ resonance, the produced quarks and gluons undergo a perturbative QCD evolution characterized by successive parton emissions and color interactions. As these emissions redistribute the energy among an increasing number of partons, the system gradually leaves the perturbative regime and enters the strongly coupled non-perturbative domain, where color confinement drives the hadronization process and converts the colored partons into color-neutral hadrons, including the $\Lambda$ and $\bar{\Lambda}$ hyperons. Recent experimental studies have shown that this transition from perturbative to non-perturbative QCD can be consistently interpreted within a quantum framework, providing new insight into the quantum nature of hadronization~\cite{Datta}. Consequently, the spin degrees of freedom of the produced hyperons are not created in isolation but originate from the same evolving QCD environment. Even after hadron formation, these spin degrees of freedom may continue to evolve through their interaction with the surrounding hadronic medium or the residual QCD environment before eventually reaching spin freeze-out.

A complementary perspective is provided by recent studies of hyperon spin dynamics in relativistic heavy-ion collisions. In particular, Sun \textit{et al.}~\cite{Sun}, in agreement with experimental observations, demonstrated that the spin degrees of freedom of the $\Lambda$ hyperons continue to evolve throughout the hadronic phase and remain dynamically coupled to the hadronic matter until the spin freeze-out temperature is reached. Their results indicate that the measured spin polarization is determined by the hadronization process and the subsequent dynamical evolution of the hadronic matter. These findings demonstrate that the spin degrees of freedom of hyperons cannot, in general, be regarded as completely isolated immediately after hadronization. In consequence, Ref.~\cite{JA} describes the spin degrees of freedom of the $\Lambda\bar{\Lambda}$ pair as a bipartite quantum subsystem coupled to a common non-perturbative QCD environment. The correlation parameter $\mu$ quantifies the degree of correlation of the common hadronization environment acting on the two spin subsystems.

Furthermore, Akamatsu \emph{et al.}~\cite{AX2} showed that open-quantum-system theory provides a consistent framework for describing the dynamical evolution of quantum states interacting with strongly coupled QCD matter. Similarly, Twagirayezu \emph{et al.}~\cite{AX1} modeled the quark--gluon plasma as a composite quantum channel, allowing decoherence, dissipation, and hadronization effects induced by the QCD medium to be incorporated into the evolution of quantum states. In this context, Ref.~\cite{JA} models the common non-perturbative QCD environment acting on the reconstructed $\Lambda\bar{\Lambda}$ spin density matrix by means of correlated Markovian and non-Markovian quantum channels. The Markovian regime describes memoryless environmental interactions, whereas the non-Markovian regime accounts for possible memory effects associated with the non-perturbative hadronization process. They reproduce the angular distributions measured by BESIII and allow us to investigate the robustness and evolution of quantum coherence and quantum correlations in the $\Lambda\bar{\Lambda}$ system.

Quantum steering is one of the quantum-correlation measures investigated for the $\Lambda\bar{\Lambda}$ system. The spin polarization and spin-correlation observables are extracted from the angular distributions of the weak decay products~\cite{JA}. These observables are then used to reconstruct the complete spin density matrix of the hyperon--antihyperon pair. Once the spin density matrix has been reconstructed, quantum steering can be evaluated directly from the resulting density matrix. Within this framework, quantum steering provides a quantitative characterization of the directional quantum correlations encoded in the experimentally reconstructed $\Lambda\bar{\Lambda}$ spin state.

The evaluation of quantum steering from experimentally reconstructed spin density matrices has recently been extended to relativistic particle systems. In particular, Afik \emph{et al.}~\cite{Afik} reconstructed the spin density matrix of top-quark pairs produced at the LHC from the measured angular distributions of their decay products and subsequently evaluated several quantum-correlation measures, including quantum steering, directly from the reconstructed density matrix. Their analysis provided experimental evidence for quantum steerability, demonstrating that experimentally reconstructed spin states provide a suitable basis for the quantitative investigation of quantum steering in high-energy particle systems.

In summary, the framework adopted in Ref.~\cite{JA} combines experimentally reconstructed spin-density matrices with correlated quantum channels to investigate quantum correlations in the $\Lambda\bar{\Lambda}$ system. Within this approach, coherence, entanglement, quantum discord, and quantum steering provide complementary information on the provide complementary information on the spin correlations of the reconstructed state and its response to the common partonic environment generated during QCD hadronization. The successful description of the BESIII spin-correlation observables supports the consistency and physical relevance of this framework for exploring quantum-information features in hyperon--antihyperon systems.

\bibliography{bib2}

\end{document}